\documentclass[journal]{IEEEtran}
\usepackage{graphicx}
\usepackage{subfigure}
\usepackage{cite}
\usepackage{amssymb, amsmath}
\usepackage{multirow}
\usepackage{epstopdf}

\begin{document}

\title{Exploiting Moving Intelligence: \\ Delay-Optimized Computation Offloading in Vehicular Fog Networks}

\author{Sheng~Zhou,~\IEEEmembership{Member,~IEEE,} Yuxuan Sun,~\IEEEmembership{Student Member,~IEEE,} Zhiyuan Jiang,~\IEEEmembership{Member,~IEEE,} Zhisheng~Niu,~\IEEEmembership{Fellow,~IEEE}
\thanks{Sheng~Zhou, Yuxuan Sun and Zhisheng~Niu are with Beijing National Research Center for Information Science and Technology, Department of
Electronic Engineering, Tsinghua University, Beijing, China.

Zhiyuan Jiang (corresponding author) is with Shanghai Institute for Advanced Communication and Data Science, Shanghai University, Shanghai, China.
}}

\maketitle

\begin{abstract}
Future vehicles will have rich computing resources to support autonomous driving and be connected by wireless technologies. Vehicular fog networks (VeFN) have thus emerged to enable computing resource sharing via computation task offloading, providing wide range of fog applications. However, the high mobility of vehicles makes it hard to guarantee the delay that accounts for both communication and computation throughout the whole task offloading procedure. In this article, we first review the state-of-the-art of task offloading in VeFN, and argue that mobility is not only an obstacle for timely computing in VeFN, but can also benefit the delay performance. We then identify machine learning and coded computing as key enabling technologies to address and exploit mobility in VeFN. Case studies are provided to illustrate how to adapt learning algorithms to fit for the dynamic environment in VeFN, and how to exploit the mobility with opportunistic computation offloading and task replication.

\end{abstract}

\section{Introduction}
To satisfy the emerging need for autonomous driving, future vehicles will not only have rich on-board sensors like cameras and radars, but also be equipped with strong computing power to process the sensing data and make driving decisions.
In addition, due to recent fatal accidents with stand-alone autonomous driving,
it becomes evident that safe and reliable autonomous driving requires effective interaction and collaboration between vehicles, and between vehicles and road side units (RSUs). Wireless technologies enable vehicle-to-vehicle (V2V) and vehicle-to-infrastructure (V2I) communications that deliver critical information, such as safety warnings and road conditions. Accordingly, vehicles can extend their sensing capability to reach blind spots, and can also jointly process the sensing data and coordinate their driving decisions. The result can be precise recognition of the environment and robust control of vehicles, leading to safer autonomous driving and more efficient road traffic.

The large number of connected vehicles, each endowed with server-level computing power, form a network with abundant amount of moving intelligence. Vehicles can contribute their computing resources, acting like fog nodes in the context of fog computing \cite{Hou2016TVT}, and thus the whole network can be regarded as vehicular fog network (VeFN). The VeFN can provide wide range of applications beyond autonomous driving. For instance, passengers can utilize the excessive computing power on their own vehicle or neighboring vehicles for computation task offloading, overcoming the device limitations as in mobile edge computing (MEC) where computing resources are co-located with base stations (BSs) \cite{mao2017mobile}. Pedestrians can also access VeFN via RSUs. To this end, VeFN combines the concepts of fog-as-a-service \cite{FA2S2017} and vehicle as the infrastructures \cite{Hou2016TVT}, and is promising in the era of artificial intelligence (AI), which calls for computing anytime and everywhere.

For autonomous driving and other computation offloading applications, \emph{delay}, accounting for both computing and transmission, is always the most demanding quality of service (QoS) requirement.
In vehicular networks, the high mobility of vehicles and the ad hoc nature of networking make timely communication and computing quite challenging.
Despite the existing research efforts on delay optimized ultra reliable V2V communications \cite{Benniseucnc}, the coupled communication and computing delays in task offloading are affected by more random factors, which is challenging to optimize.
High mobility introduces difficulties in jointly adapting wireless and computing resources with respect to the time varying system conditions. Moreover, the adaptation requires fresh system state information of channels and computing power, which is unfortunately hard to obtain.


Nevertheless, mobility is not always an obstacle.
As shown in the premier work by Grossglauser and Tse, mobility can increase the capacity of wireless ad hoc networks by increasing the probability of making contacts between nodes \cite{TseMobility2002}. Moreover as shown in \cite{Dhillonwcl} \cite{Dhillontvt}, mobility can increase the successful downloading probability of files in caching systems, with more chances for end users to experience good channels and file holders.
We believe that mobility is also beneficial to the task offloading in VeFN.
For example, the probability that vehicles with excessive computing resources appear in the vicinity of an end user can increase with the speed of vehicles \cite{Jiang2017IoT}. In this context, how to guarantee offloading delay while at the same time exploiting the \emph{diversity} brought by mobility becomes an intriguing research issue.

In this article, we first introduce the concept of VeFN with the corresponding delay requirement, and review state-of-the-art for cloud computing under dynamic conditions in vehicular networks.
	We propose the architecture of VeFN with three major offloading modes in the next section, and analyze the advantages and challenges of these modes.
	We then discuss why mobility is both a foe and a friend of computation offloading in VeFN, and propose two key solutions, i.e., learning while offloading and coded computing, to optimize the offloading delay. We carry out two case studies that address and exploit the mobility in VeFN, respectively.
	In the first case study, the existing learning algorithms are revisited and modified so that they can adapt to the varying network topologies and workloads. Coded computing is combined with learning techniques to further improve the service reliability.
	In the second case study, the optimal task replication policy is derived, providing insights that balanced assignment optimizes the delay performance.
	Finally we conclude the article with outlook on future research directions.

\section{Vehicular Fog Networks: Architecture and State-of-the-Art} \label{sec_architecture}

\begin{figure*}[!t]
	\centering
	\includegraphics[width=0.6\textwidth]{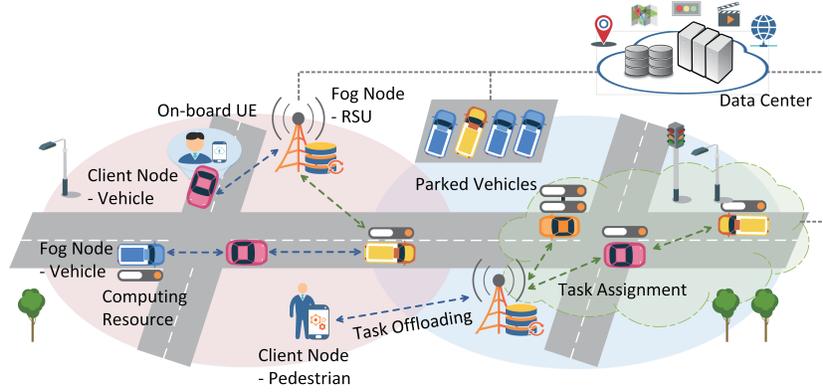}
	\caption{Illustration of task offloading in VeFN.}
	\label{fig_arch}
\end{figure*}

The VeFN integrates the computing resources of vehicles and RSUs, and provides diverse fog computing services and applications for vehicles and mobile users.
As shown in Fig. \ref{fig_arch}, RSUs and vehicles (moving and parked), that can provide computing resources, are regarded as fog nodes. Computation tasks with different workloads and delay requirements are generated by the client nodes, including vehicles requiring excessive computing supports, on-board user equipments (UEs), pedestrians, etc. These tasks are offloaded from the client nodes to the VeFN for processing.
Note that each vehicle can act as either a fog node or a client node, denoted by \emph{fog vehicle} and \emph{client vehicle}, respectively. The role of each vehicle can change over time, depending on whether it has surplus computing resources to contribute to the network, or whether it requires support from other nodes for task offloading.

\begin{table*} [!t]
	\caption{Typical applications in VeFN and the corresponding delay requirements}
	\label{applications}
	\centering
	\begin{tabular}{|c|c|c|}
		\hline
		 \textbf{Application} &\textbf{Type}& \textbf{Delay Requirements} \\
		 \hline
		Cooperative collision avoidance  &Safety& Delay bound, 10ms\\
		\hline
		Vehicle platooning &Safety& Delay bound, 25ms \\
		\hline
		Collective perception of environment & Safety& Delay bound, 500ms \\
		\hline
		Vehicle scheduling  & Non-safety& Average delay, 1s \\
		\hline
		Virtual reality and augmented reality & Entertainment& Delay bound, 10ms\\
		\hline
		Cloud gaming & Entertainment & Average delay, 100ms$\sim$1s\\
		\hline
		Road monitoring and flow optimization  & IoT & Average delay, seconds$\sim$minutes\\
		\hline
	\end{tabular}
\end{table*}

Some typical applications in VeFN and their corresponding delay requirements are summarized in TABLE \ref{applications}, where the data is from \cite{mao2017mobile} and 3GPP TR 22.886 \cite{3gpp22886}.
The key performance metric of task offloading is \emph{delay}, consisting of three parts \cite{mao2017mobile}: uploading delay related to the input data size of a task that needs to be processed, computing delay at the fog node which is related to the computational complexity and the input data size, and downloading delay related to the output data size.
All these delays are affected by the communication bandwidth used to transmit the data, and the computing power to process the task at the fog nodes.
For delay-critical applications, the hard delay bound represents the longest allowable delay that cannot be violated. For delay-tolerant applications, tasks do not have an exact deadline, but timely feedbacks are still favorable.
Such applications include traffic flow optimization, entertainments and IoT applications, and average offloading delay can be used as the performance metric.

\subsection{Architecture and Offloading Modes in VeFN}

The scattered computing resources in VeFN bring a variety of offloading routes in the VeFN. To support the data transmissions between client nodes and fog nodes, multiple communication techniques are jointly used, including IEEE 802.11p-based dedicated short-range communications (DSRC) and LTE-V, which enable vehicle-to-everything (V2X) communications such as V2V, V2I and vehicle-to-pedestrian (V2P) communications. Pedestrians get access to the RSUs through 3G or 4G LTE, and on-board UEs can offload tasks to the vehicle they ride via Bluetooth. As shown in Fig. \ref{fig_arch}, the computation task offloading in VeFN is classified into three major modes:

\begin{itemize}

\item \emph{Vehicle-Vehicle Offloading}: Vehicles can directly offload their tasks (including the tasks offloaded by their passenger UEs) to the neighboring fog vehicles. In this case, each client vehicle first discovers the available fog vehicles in its communication range. To keep a relatively long contact time, the moving directions and velocities should be considered, which can be acquired by V2X communication protocols. Multiple fog vehicles may be available at the same time. Offloading decisions about which fog vehicles to select are made by client vehicles independently in a distributed manner, since it is difficult to acquire global information about the vicinity area, and there might not be a centralized entity to make such decisions.

\item \emph{Vehicle-RSU-Vehicle Offloading}: When the surrounding fog vehicles can not satisfy the computing needs of client vehicles, tasks are offloaded to nearby RSUs. RSUs may compute the tasks by their own computing resources, or further assign the tasks to other fog vehicles without direct wireless connection to the client vehicles. RSUs are able to master more information about communication bandwidth and computing resources. Hence centralized task assignment can optimize the utilization of computing resources and the QoS. Computing results are finally transmitted back to the client vehicle via its associated RSU.

\item \emph{Pedestrian-RSU-Vehicle Offloading}: The contact durations of pedestrians and vehicles are often very short and the connectivities are quite unstable. Therefore, in VeFN, RSUs can first collect the computation tasks from pedestrians. Then the tasks are handled by themselves, or offloaded to the fog vehicles. The computing results can be fed back by the fog vehicle to its nearby RSU, and then delivered back to the original RSU via the backhaul, if the fog vehicle has already moved away from the original RSU.

\end{itemize}

\subsection{State-of-the-Art on Computing in Vehicular Networks}

There are some recent efforts focusing on the resource management of communication and computation in the context of VeFN. 
Researchers start from evaluating the feasibility of employing vehicles as fog nodes.
Hou \emph{et al.}\cite{Hou2016TVT} analyzed communication and computing capacity of vehicles using real traces of vehicles in Beijing and Shanghai. Simulation results show that the communication and computing resources of both parked and moving vehicles have great potential to enhance the static fog computing network.

For flexible computing resource management, a software-defined vehicular network architecture was proposed by Choo \emph{et al.} \cite{choo2017sdvc}, in which a centralized vehicular cloud (VC) controller periodically collects the mobility and resource status of fog vehicles, estimates their instantaneous locations and computation loads upon task requests, and allocates the computing and bandwidth resources for each task.
In terms of resource allocation schemes, Zheng \emph{et al.} \cite{zheng2015smdp} considered a dynamic VC consisted of moving vehicles. The arrival and departure of vehicles follow the Poisson process, and each vehicle is equipped with equal computing power. Tasks are collected by a central controller, and assigned to fog vehicles to maximize the average utility related to delay, energy consumption and resource occupation.
However, the basis of adopting centralized schemes is holding the instantaneous state information of the whole system, which may lead to high signaling overhead, and is thus hard to implement in real systems.

Task offloading decisions can also be made by each client node independently in a distributed manner, corresponding to the vehicle-vehicle offloading mode.
Feng \emph{et al.} \cite{feng2017TVT} proposed a distributed VeFN architecture, where RSUs and vehicles can offload tasks to their neighboring nodes based on their distributed decisions.
They designed a task offloading algorithm based on ant colony optimization, in order to maximize the sum utility of offloaded tasks related to delay, and evaluate it in a system level simulator using real traces.
However, the RSUs and vehicles are treated equally in \cite{feng2017TVT}, while in real systems, RSUs often know more about the network conditions, which should be more effectively exploited.

\section{Mobility: Foe and Friend for the Offloading Delay} \label{sec_mobility}

The \emph{mobility} of vehicles make the VeFN highly dynamic and volatile, which bring both challenges and opportunities for computation task offloading. In this section, we interpret the intuition about why mobility is acting both as a foe and a friend for timely computing, and identify potential ways to address and exploit the mobility.

\subsection{Mobility as a Foe}
Mobility brings more randomness and uncertainties to the delay performance of the offloaded tasks. First, the VeFN can be viewed as an intermittently connected wireless network, in the sense that the network topology changes over time, and connection durations of V2X, including V2V, V2I and V2P, are quite limited. This inherently stems from the physical limitations of the range of wireless communications and the high mobility of vehicles, and significantly limits the effectiveness of VeFN.
Second, the wireless channel states and thus the interference between V2X vary fast across time, depending on many factors such as relative speed, neighboring vehicles' transmit power and surrounding scatters, and they are hard to model or to predict.
Third, similar to the MEC systems, the computation tasks are generated randomly with different delay requirements and workloads, and the computing power of RSUs and vehicles varies, producing highly dynamic and non-uniform computation loads.
These factors bring challenges to collect information and to make optimal offloading decisions and resource allocation in a timely manner, which is critical for those safety-related applications with hard delay bounds.

\subsection{Mobility as a Friend}
However, the limitation due to intermittent connectivity can be elevated by, somewhat surprisingly, vehicle mobility. This can be illustrated by the following three factors:

\begin{itemize}
	\item
	\emph{Mobility increases the probability of making contact}: As shown by several existing work such as \cite{TseMobility2002}, the mobility of nodes in an intermittently connected network can be beneficial since mobility creates more chances of contacts between nodes, and thus the probability of communication and task offloading between the nodes also increases.
	\item
	\emph{Mobility decreases contact time interval}: Many applications of VeFN rely on consecutive contacts between V2V and V2I, such that the inputs and outputs of tasks can be communicated separately. Therefore, the offloading delay is related to the contact time interval significantly. In this regard, vehicle mobility decreases the time interval and hence reduces the offloading delay.
	\item
	\emph{Predictable Mobility}: Despite the high mobility of vehicles, their trajectories are limited to roads, and their speeds are highly related to the traffic conditions. Thus one can predict the mobility of vehicles to certain extent and carry out prediction-based task offloading.
\end{itemize}

\subsection{Address and Exploit the Mobility}

To release the aforementioned potentials brought by mobility, while reduce the time overhead for information collection and online decisions, we resort to machine learning approaches to track the system dynamics. Moreover, we identify a set of solutions that falls into the concept of coded computing, which can enhance the reliability of computing with efficient resource utilization.

\textbf{1) Learning while offloading}

Because both communication and computation environments depend on many factors and vary fast in VeFN, the offloading delay is very complex to model and to predict, especially for distributed task offloading.
Instead of acquiring all the related state information to infer the offloading delay of candidate fog nodes before making the offloading decision, client nodes can try different candidates by offloading several tasks and observing the delay on the go. In other words, the environment is learnt while tasks are being offloaded.
Such learning algorithms should have low complexity and fast convergence to track the dynamic environment. They should also effectively balance the so-called \emph{exploration-exploitation tradeoff}: to explore more and get more accurate estimations about candidate fog vehicles, or to select the empirically best fog vehicle, hoping to minimize the instantaneous offloading delay. In the context of learning, the objective is to minimize the \emph{regret}, that is, the performance loss of learning algorithms compared with the genie-aided optimal solution.

Multi-armed bandit (MAB) \cite{auer2002finite} is a promising method to perform learning on the go. 
In classical MAB, a decision maker faces a fixed number of candidate actions, whose rewards are governed by different distributions that are unknown in prior. The decision maker tries one action at a time, observes the reward, and gradually learns the performance of different candidates while minimizing the regret.

However, existing MAB algorithms can not be applied in VeFN directly, since the network topologies are not fixed, with neighboring fog vehicles coming and leaving unexpectedly. In addition, the workloads of tasks are varying over time, but such variations have not been considered in existing MAB problems. We revise MAB-based learning algorithm to address the dynamic topology and task workloads in vehicle-to-vehicle offloading, which will be illustrated in the first case study.

To exploit the mobility, supervised learning methods can also be adopted to learn the mobility of vehicles and predict their speeds and trajectories. Then both computing and bandwidth resources of fog vehicles can be better allocated. For example, vehicles can predict which neighboring vehicles may have longer contact duration, and RSUs can forecast the occurrence of handover and proactively fetch the computing data or migrate some computing services of client vehicles.

\textbf{2) Coded computing}

\begin{figure*}[!t]
	\centering
	\includegraphics[width=0.6\textwidth]{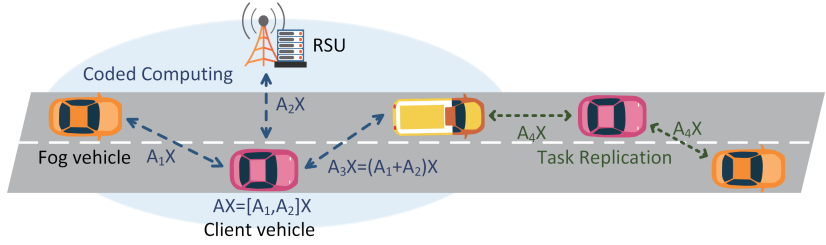}
	\caption{Illustration of coded computing and task replication in VeFN.}
	\label{fig_coded}
\end{figure*}

As a foe, mobility makes the computing services at each fog node unreliable.
But as a friend, mobility also brings opportunities for client nodes to meet more fog nodes. Equivalently, the computing resources of VeFN become richer. To exchange the redundancy of computing resources for reliability, \emph{coded computing} serves as an efficient tool. Consider a $(n,m)$ maximum distance separable (MDS) coding scheme. If $n$ fog nodes can provide computing services for a client node, a task can be decomposed into $m$ subtasks with $m\leq n$ using the minimum latency coding technique \cite{li2017coding}, encoded into $n$ coded tasks, and then offloaded to these $n$ fog nodes. Once the earliest $m$ computing results are successfully delivered back, the task is completed.
An example of coded computing in VeFN is shown in Fig.~\ref{fig_coded}. Three fog nodes (two fog vehicles and one RSU) can serve the client vehicle. 
The input data of a matrix multiplication task is decomposed as two submatrices, and encoded to be three subtasks.
The original task is successfully computed if two of the three fog nodes complete their subtasks.
As a result, fog nodes can cooperate with each other to better utilize their computing resources,
and the uncertainties brought by mobility, such as intermittent connectivity, can be addressed.
Coded computing can also be applied to share the computing resources of fog nodes for \emph{multiple} client vehicles. 



Coded computing can actually cover a large variety of mappings between computing resources and tasks, among which a special case is \emph{task replication}, using the simplest repetition coding. Each task is directly offloaded to multiple fog nodes simultaneously and processed independently. If one of the selected fog nodes completes the task before the deadline, it is successfully executed.
It is crucial to balance the reliability gain with more replications and the resource occupation alongside, and the offloading opportunities require non-trivial allocation among multiple clients.
Our key contribution is that, we derive the optimal task replication decisions for multiple clients in VeFN in the second case study, providing insights that the balanced task assignment optimize the delay performance.
We also discuss how to combine MAB and coded computing in the first case study, and show the performance gain through coding.



\section{Case Studies}
In this section, we carry out two case studies that apply the aforementioned learning and coded computing methods, proving their great potentials for timely computing in VeFN.

\subsection{Learning-based Task Offloading in VeFN}
To guide the task offloading in VeFN by using learning-based method, we first focus on a distributed scenario with vehicle-vehicle offloading. Consider a client vehicle of interests who generates computation tasks in sequence, and it makes the offloading decisions on which fog vehicle to handle each task, to minimize the average offloading delay.
%

Each client vehicle is hard to acquire the information about available computing resources and channel states for its own tasks, thus it has no idea which fog vehicle performs the best when making task offloading decisions.
It has to learn the average delay of each candidate fog vehicles based on observations of the offloading delay associated to each candidate vehicle.

As mentioned, MAB-based learning method can be used to design the task offloading algorithm, yet still requires adaptations to fit for the dynamic environment in VeFN.
To make MAB effective in VeFN, we redesign the utility function of conventional MAB by considering the following three key factors:
1) The empirical delay of the offloaded tasks, which is the delay performance of the candidate fog vehicles one has learnt. 2) The appearance time of each fog vehicle. The client vehicle should try more to the newly appeared fog vehicles, while exploit what it has already learnt about the existing fog vehicles.
3) The workload of each task. Since the offloading delay is proportional to the workload, intuitively the client vehicle should try to exploit different fog vehicles when the workload is low, so that the regret due to learning can be reduced, and vice versa.
We propose an Adaptive Learning-based Task Offloading (ALTO) algorithm, proving that the complexity of the ALTO algorithm is linear with the number of candidate fog vehicles, and the regret grows sublinearly over time \cite{Sun2018ICC}.

For simulations, we download the map of a 12km stretch of G6 Highway in Beijing from Open Street Map (OSM)
and generate the traffic by Simulation of Urban Mobility (SUMO).
Fog vehicles are equipped with heterogeneous computing power, with CPU frequencies in the range of $[2, 5]\mathrm{GHz}$. The input data size of each task is uniformly distributed in $[0.2, 1]\mathrm{Mbits}$. We assume that tasks are of equal computation intensity $1000 \mathrm{cycles/bit}$ and the size of the output data is neglected.
The wireless connectivity is intermittent, with the probability of successful transmission $p=0.9$ or $0.95$.

\begin{figure}[!t]
	\centering
	   \subfigure[Avearge delay of ALTO algorithm.]{\label{mab_fig1}	
		\includegraphics[width=0.45\textwidth]{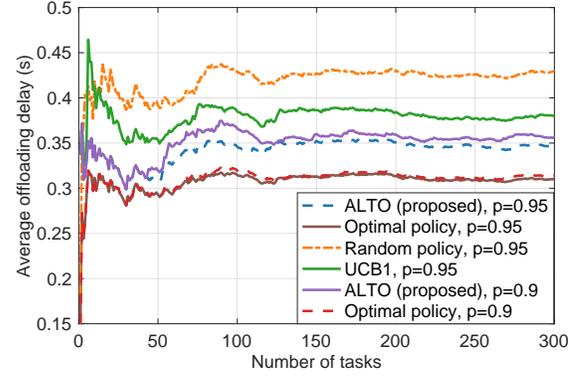}}
	\subfigure[Task completion ratio.]{\label{mab_fig2}			
		\includegraphics[width=0.45\textwidth]{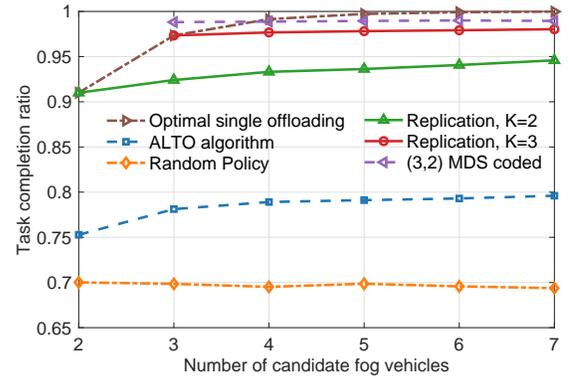}}		
	\caption{Delay performance of MAB-based task offloading algorithms.}
	\label{Performance_SUMO}
\end{figure}

As shown in Fig. \ref{mab_fig1}, the proposed ALTO algorithm is compared with three baselines: \textbf{UCB1} is the conventional MAB-based learning algorithms \cite{auer2002finite}. \textbf{Random Policy} is a naive policy in which the client node randomly selects a fog vehicle for each task. \textbf{Optimal Policy} is a genie-aided one and always selects the best fog vehicle with minimum offloading delay. Since fog vehicles may appear as candidates or leave, the average delay fluctuates over time. According to the simulations, UCB1 does not work well with moving vehicles and time-varying workloads, while our proposed ALTO algorithm performs better in terms of average offloading delay. This highlights the importance of revisiting machine learning methods to deal with the dynamics in VeFN.

To further improve the QoS, we integrate learning with task replication and (3,2) MDS coding.
Still, the global state information is unknown and needs to be learnt, and intermittent connectivity is considered with $p=0.9$.
Fig. \ref{mab_fig2} observes the ratio of tasks that are completed before a deadline $0.55\mathrm{s}$, 
and $K$ is the number of replications.
The optimal single offloading policy is the same as the Optimal Policy in Fig. \ref{mab_fig1}.
It is shown that compared with ALTO algorithm with single offloading, the service reliability is substantially improved through task replication, while light replication with $K=2$ or $K=3$ provides most of the gains.
Meanwhile, the task completion ratio of MDS coding reaches over $98\%$ with small number of fog vehicles, and even outperforms the optimal single offloading policy. This is because MDS coding reduces the workload of each coded subtask, and can further exploit the computing resources of multiple fog vehicles.

\subsection{Delay-Constrained Task Replication Exploiting Vehicle Mobility}

\begin{figure}[!t]
	\centering			
		\includegraphics[width=0.45\textwidth]{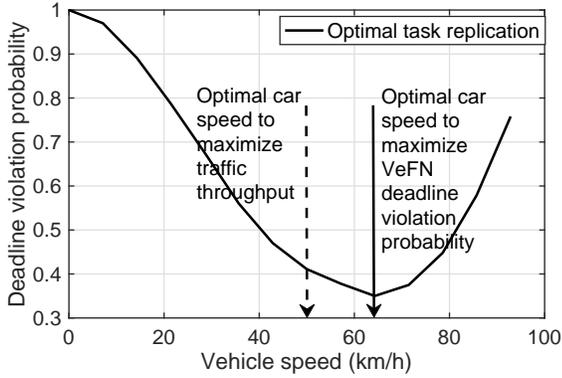}
	\caption{Task replication and corresponding performance analysis for the deadline violation probability in VeFN.}
	\label{rep_fig}
\end{figure}

In this case study, we focus on the vehicle/pedestrian-RSU-vehicle offloading mode with task replication.
Each RSU collects tasks from pedestrians or vehicles, and then assigns them to the fog vehicles coming into its coverage.
Tasks have hard delay bound that can not be violated.
Multiple tasks collected by each RSU waiting to be executed form a task queue. Each fog vehicle is assigned one task at a time, and the task replication is used, that is, each task can be assigned to multiple fog vehicles and executed independently.

Our objective is to minimize the deadline violation ratio of tasks, by deciding which task should be allocated to which fog vehicle. Assume that the arrival of fog vehicles follows Poisson process, and the sojourn time (the duration of task assignment and result feedback) of each task at each fog vehicle follows exponential distribution with homogeneous exponent. This enables a finite horizon Markov decision process (MDP) formulation of the problem, and we derive the optimal policy called balanced task assignment (BETA)  \cite{Jiang2017IoT}. The main intuition of BETA is that unfinished tasks with least number of offloading replications should be scheduled first when a new fog vehicle arrives, and this \emph{balanced} allocation of computing resources is optimal and avoids unnecessary service wastes.

We further investigate how mobility affects the computing performance. Under linear speed-density relationship which is widely used in vehicle traffic theory, the optimal vehicle speed that maximizes the traffic throughput (the number of vehicles that pass through the road) is $\frac{V_{max}}{2}$, where $V_{max}$ is the maximum allowed speed on the road. Generalizations to non-linear speed-density relationship can be found in \cite{Jiang2017IoT} while the linear model is sufficient to capture the essence. In the VeFN system, the optimal vehicle speed that minimizes the deadline violation ratio is proved to be $\frac{2V_{max}}{3}$, meaning that when vehicles move faster within $\frac{2V_{max}}{3}$, the reliability of computing services increases \cite{Jiang2017IoT}. Note that these results are from a statistical point of view, i.e., the optimal speeds are averaged among all vehicles on the road.
As shown in Fig. \ref{rep_fig}, when the vehicle speed increases, the deadline violation probability first decreases and then goes up.
This is mainly because the mobility brings more opportunities for the RSUs to meet fog vehicles, and thus the reliability first increases. However, as the vehicle speed rises, the density of vehicles finally becomes too low to support the task requirements.

\section{Conclusion and Outlook}

In this article, we have presented the VeFN concept with latest literature review, and discussed the role of mobility for timely computing in VeFN as a foe and as a friend. Enabling technologies to address and exploit mobility, including machine learning and coded computing, are introduced and their initial adoptions in VeFN are illustrated through two case studies. Whilst notable gains in terms of lower average delay and lower delay-bound violation probability are proved via MAB-based learning scheme and task replication, more efforts are needed to truly realize the potentials of VeFN accompanied by mobility.

First, computation task partition plays an important role in task offloading, yet it has been rarely touched.
Considering the heterogeneity of computing resources, task partition helps to optimize the utilization of resources and balance the workloads.

Second, mobility prediction, either model-based or reinforcement learning-based, can be exploited to reduce offloading delay, by proactive resource provisioning or computation pre-fetching. Accordingly, coded computing over tasks generated at different time is also worth investigating.

Last but not least, conventional encryption and authentication schemes may be too slow to perform in VeFN with high dynamics, especially for delay-critical applications.
Guaranteeing security and privacy in task offloading call for novel designs, and may trigger a new dimension to understand the mobility.

\section*{Acknowledgement}
This work is sponsored in part by the Nature Science Foundation of China (No. 61871254, No. 91638204, No. 61571265, No. 61861136003, No. 61621091), National Key R\&D Program of China 2018YFB0105005, and Intel Collaborative Research Institute for Intelligent and Automated Connected Vehicles.


\begin{thebibliography}{99}



\bibitem{Hou2016TVT}
X. Hou, Y. Li, M. Chen, D. Wu, D. Jin, and S. Chen, ``Vehicular fog computing: A viewpoint of vehicles as the infrastructures," \emph{IEEE Trans. Veh. Tech.}, vol. 65, pp. 3860-3873, Jun. 2016.

\bibitem{mao2017mobile}
Y.~Mao, C.~You, J.~Zhang, K.~Huang, and K. B. Letaief, ``A survey on mobile edge computing: The communication perspective,'' \emph{IEEE Commun. Surveys Tut.}, vol. 19, no. 4, pp. 2322-2358, Fourthquarter 2017.

\bibitem{FA2S2017}
N. Chen, Y. Yang, T. Zhang, M. T. Zhou, X. L. Luo, and J. Zao, ``Fog as a Service Technology,'' \emph{IEEE Commun. Mag.}, vol. 56, no. 11, pp. 95-101, Nov. 2018.


\bibitem{Benniseucnc}
M. I. Ashraf, Chen-Feng Liu, M. Bennis, and W. Saad, ``Towards low-latency and ultra-reliable vehicle-to-vehicle communication," \emph{European Conf. Netw. Commun. (EuCNC)}, Oulu, Finland, 2017.

\bibitem{TseMobility2002}
M.~Grossglauser, and D. N. C. Tse, ``Mobility increases the capacity of ad hoc wireless network," \emph{IEEE/ACM Trans. Netw.}, vol.~10, no.~4, pp.~477-486, Mar. 2002.

\bibitem{Dhillonwcl}
S. Krishnan, and H. S. Dhillon, ``Effect of user mobility on the performance of device-to-device
networks with distributed caching,'' \emph{IEEE Wireless Commun. Lett.}, vol. 6, no. 2, pp. 194-197,
Apr. 2017.

\bibitem{Dhillontvt}
S. Krishnan, M. Afshang, and H. S. Dhillon, ``Effect of retransmissions on optimal caching in cache-enabled small cell networks,'' \emph{IEEE Trans. Veh. Technol.}, vol. 66, no. 12, pp. 11383-11387,
Dec. 2017.

\bibitem{3gpp22886}
3GPP TR 22.886, ``Study on enhancement of 3GPP support for 5G V2X services," V15.1.0, Mar. 2017.


\bibitem{Jiang2017IoT}
Z.~Jiang, S.~Zhou, X.~Guo, and Z.~Niu, ``Task replication for deadline-constrained vehicular cloud computing: Optimal policy, performance analysis and implications on road traffic," \emph{IEEE Internet Things J.}, vol. 5, no. 1, pp. 93-107, Feb. 2018.




%


\bibitem{choo2017sdvc}
J. S. Choo, M. Kim, S. Pack, and G. Dan, ``The software-defined vehicular cloud: A new level of sharing the road," \emph{IEEE Veh. Technol. Mag.}, vol. 12, no. 2, pp. 78-88, Jun. 2017.

\bibitem{zheng2015smdp}
K. Zheng, H. Meng, P. Chatzimisios, L. Lei, and X. Shen, ``An \protect{SMDP}-based resource allocation in vehicular cloud computing systems," \emph{IEEE Trans. Ind. Electron.}, vol. 62, no. 12, pp. 7920-7928, Dec. 2015.



\bibitem{feng2017TVT}
J. Feng, Z. Liu, C. Wu, and Y. Ji, ``\protect{AVE}: Autonomous vehicular edge computing framework with \protect{ACO}-based scheduling," \emph{IEEE Trans. Veh. Technol.}, vol. 66, no. 12, pp. 10660-10675, Dec. 2017.

%

\bibitem{auer2002finite}
P.~Auer, N.~Cesa-Bianchi, and P.~Fischer, ``Finite-time analysis of the multiarmed bandit problem,'' \emph{Machine learning}, vol.~47, no. 2-3, pp.
235--256, May 2002.

\bibitem{li2017coding}
S. Li, M. A. Maddah-Ali, and A. S. Avestimehr, ``Coding for Distributed Fog Computing," \emph{IEEE Commun. Mag.}, vol. 55, no. 4, pp. 34-40, Apr. 2017.



\bibitem{Sun2018ICC}
Y.~Sun, X.~Guo, S.~Zhou, Z. Jiang, X.~Liu, and Z.~Niu, ``Learning-based task offloading for vehicular cloud computing systems," \emph{IEEE Int. Conf. Commun. (ICC)}, Kansas City, MO, USA, May 2018.








%


\end{thebibliography}
\end{document}